\documentclass[pre,floatfix,showpacs]{revtex4}

\usepackage{amssymb}
\usepackage{graphicx}

\begin{document}




\title{A new dimension to Turing patterns}

\author{Teemu Lepp\"anen} 
\author{Mikko Karttunen} 
\author{Kimmo Kaski}
\affiliation{Research Centre for Computational Science and Engineering, Helsinki University of Technology,
P.O. Box 9400, FIN-02015 HUT, Finland}
\author{Rafael A. Barrio}
\affiliation{Instituto de F\'\i sica, Universidad Nacional Aut\'onoma de M\'exico (UNAM)
Apartado Postal 20-364, 01000, M\'exico D.F., MEXICO}
\author{Limei Zhang}
\affiliation{Departamento de Fisiolog\'{\i}a Celular, Facultad de Medicina,
Universidad  Nacional Aut\'onoma de M\'exico,  04510, M\'exico D.F., MEXICO}


\begin{abstract} 

It is well known that simple reaction-diffusion systems can display 
very rich pattern formation behavior. Here we have studied two examples of such systems
in three dimensions. First we investigate the morphology and stability 
of a generic Turing system in three dimensions and then the 
well-known Gray-Scott model. In the latter case, we added a small number of morphogen sources
in the system in order to study its robustness and the formation of 
connections between the sources. Our results raise the question of whether 
Turing patterning can produce an inductive signaling mechanism for neuronal growth. 

\end{abstract} 



\maketitle


\section{Introduction}

Fifty years ago Alan Turing proposed the now famous reaction-diffusion
system involving two chemicals \cite{turing}  to model biological pattern formation,
and ``morphogenesis''. Since then, it has been extensively used in studying various specific 
problems in mathematical biology \cite{murray}, and currently there exists a vast 
literature on the subject, see e.g. Ref.~\cite{maini1} and references therein. 
In Turing systems nonoscillatory (in time) bifurcations occur provided that there is a substantial
difference in the diffusion rates of the two substances. The appearance of a 
spatial pattern from the destabilization of  a homogeneous state in the absence 
of diffusion is known as diffusion-driven instability.

The validity of the Turing models for describing developmental, ecological and chemical processes 
has been a matter of debate due to the difficulty of finding and identifying the morphogens,
and it was not before 1989 that  the first observation of a non-oscillatory steady 
pattern was found in a real chemical reaction \cite{cima}. More recently, it was 
pointed out by Kondo and Asai \cite{kondo} that the stripes on the skin of the fish 
{\it Pomacanthus Imperator} had the characteristics of a Turing pattern, and this was
verified later by two-dimensional numerical studies of growing domains \cite{varea}. 

The real problem with Turing models is that bifurcation analysis and numerical calculations 
in one and two dimensions tend to produce very similar patterns, for a wide variety of
models. A detailed study of patterns found in two dimensions \cite{barrio1}  
revealed that a cubic term favors striped patterns, while a quadratic term produces spots, 
the latter being more robust. However, it is possible to produce more complicated shapes by 
coupling a Turing system to another Turing system \cite{barrio1}, to a mechanochemical model \cite{shaw}, 
or to a chemotactic mechanism \cite{painter}. The resulting patterns bare striking resemblance to the 
ones found in the corresponding biological systems. As it is not easy to
experimentally set up the stringent conditions for the appearance of a Turing 
instability, namely, spatial homogeneity and a very large ratio of the diffusion 
constants, there has been a lot of research towards investigating the effect on 
patterns due to selection of having an inhomogeneous domain \cite{maini2}, different  
boundary conditions \cite{barrio2}, the presence of possible sources of 
morphogens \cite{barrio3}, growing domains \cite{varea}, and  
curvature  \cite{varea2,faustino}.  
 
Recently there has been a suggestion that the concepts of positional information 
and chemical signaling present in the Turing systems could play an important role 
as an inductive mechanism for neuronal connections \cite{olivares}. We are particularly 
interested in pursuing this idea, but any specific 
mechanism for Turing patterning in a real tissue has to happen in 
three dimensions. Since there have been relatively few systematic studies of three-dimensional 
patterns, we have recently done some preliminary numerical calculations 
in three dimensions \cite{teemu}. The results are analogous to the patterns in two dimensions, 
although it is not obvious that the pattern selection should be the same as in two dimensions. 

In this paper we present a series of numerical 
calculations carried out in a cube with periodic boundary conditions. 
We have chosen two simple generic models for this purpose: a well known version of 
the Gray-Scott model \cite{xmorphia}, and a general Turing 
model analyzed by Barrio {\it et al.} \cite{barrio1} that 
allows the investigation of quadratic and cubic terms separately. 
 
\section{The Models}  

In this paper we are going to study these two models mainly numerically in three dimensions.
In its general form the Turing system for modelling the evolution of the concentrations of two chemicals 
is given as
\begin{eqnarray}
\partial_t U & = & D_U \nabla^2 U + f(U,V)\nonumber \\
\partial_t V & = & D_V \nabla^2 V + g(U,V) 
\label{turing}
\end{eqnarray}
where $U \equiv U(\vec{x},t)$ and  $V \equiv V(\vec{x},t)$ are the concentrations, and
$D_U$ and $D_V$ the respective diffusion constants. The reactions are modeled by the functions
$f$ and $g$ which are typically non-linear.

As has been pointed out earlier \cite{barrio1} in relation to the studies of two 
dimensional systems, there may exist a number of different admissible modes 
having the same wave number. In three dimensions, as studied here, this degeneracy
occurs already at small wave numbers and the question of which mode will dominate the
evolution becomes even more difficult to answer. This is obvious when one recalls that in a
finite grid the admissible modes are not continuous but discrete, i.e., 
\begin{equation}
|\vec{k}|= 2 \pi \sqrt{\left(\frac{n_x}{L_x}\right)^2 +\left(\frac{n_y}{L_y}\right)^2 
+ \left(\frac{n_z}{L_z}\right)^2},
\label{eq:wavevector}
\end{equation}
where $L_{x/y/z}$ denote the system size in respective directions and $n_{x/y/z}$
the respective wave number indices.

\subsection{General Turing system}

As our first model, we studied a general Turing system 
proposed by Barrio {\it et al.} \cite{barrio1}.
The equations of motion are obtained by Taylor expansion around the stationary uniform
solution $(U_c,V_c)$. Keeping terms up to cubic order, the equations of motion read as follows
\begin{eqnarray}
\partial_t u & = & D \delta \nabla^2 u + \alpha u(1-r_1 v^2)  + v(1-r_2 u) \nonumber \\
\partial_t v & = & \delta \nabla^2 v + v(\beta + \alpha r_1 uv) + u(\gamma + r_2 v), 
\label{eq:barrio}
\end{eqnarray}
where $u=U-U_{c}$ and $v=V-V_{c}$ making the point $(u,v) = (0,0)$ a stationary solution.
The constant $\delta$ is a scaling factor and $D$ is the ratio between the 
diffusion constants of the two chemicals. It is important to notice that
$D \ne 1$ is required for the diffusion-driven instability to occur.
The parameters $r_1$ and 
$r_2$ of the non-linear interactions correspond to, as can be verified by symmetry arguments,
stripes and spot patterns, respectively.  

The linear analysis for this system was done by  Barrio {\it et al.} \cite{barrio1}. 
For completeness, the results that are relevant for this study are 
summarized here. The system described 
by Eq.~(\ref{eq:barrio}) has a stationary solution for $(u,v) = (0,0)$ and a second solution for 
$$
v=\frac{-(\alpha + \gamma)}{\beta+1}u.
$$
By setting $\alpha = - \gamma$ it is possible to enforce  $(0,0)$ as the only 
stationary solution. The dispersion relation for the linearized equation can be found 
by noticing that the spatial variation of $u$ and $v$ can be given in a plane wave form, 
and in order to find the eigenvalues standard Floquet analysis with $u=u_0 \exp (\lambda t)$
and  $v=v_0 \exp (\lambda t)$ is used. This leads to 
\begin{equation}
\lambda^2 - B \lambda + C=0
\label{eq:quadr}
\end{equation}
where $B=(\alpha +\beta - \delta k^2 (1+D))$, 
$C=(\alpha -\delta D k^2)(\beta - \delta k^2) + \alpha$, and 
$k^2 = \vec{k} \cdot \vec{k}$. 

\begin{figure}[!]
\begin{center}
\includegraphics[width=.8\textwidth]{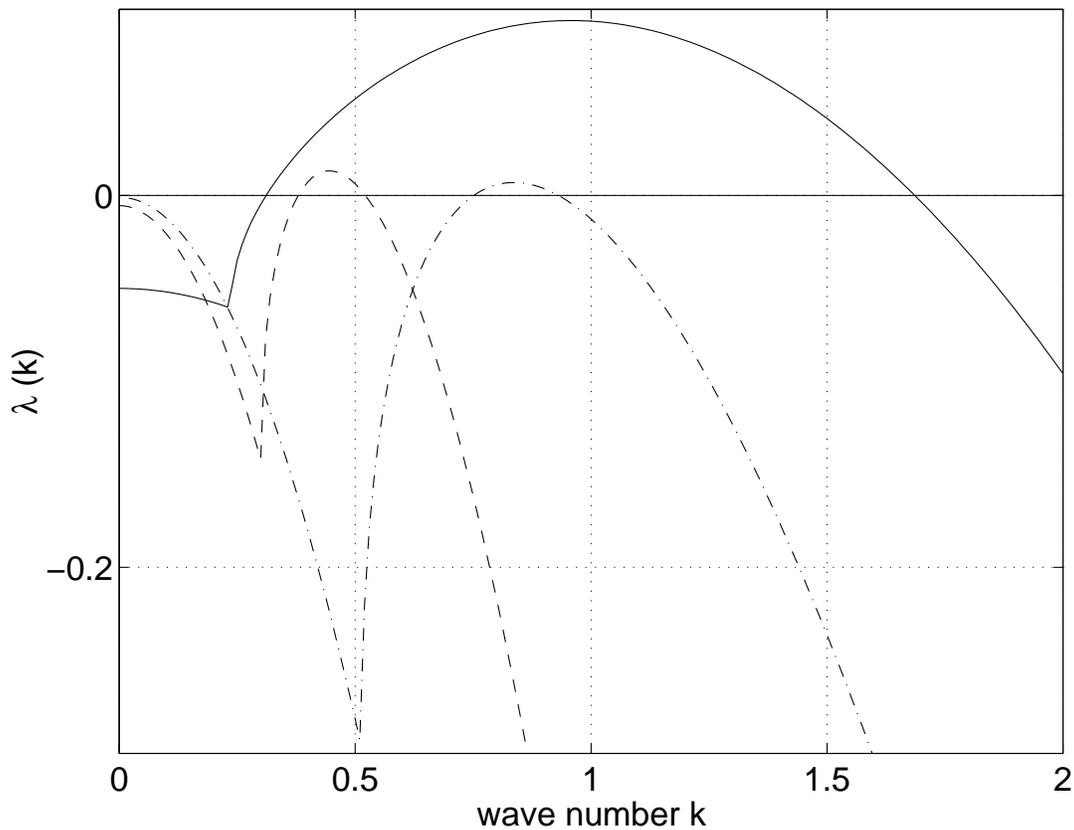} 
\caption{
The dispersion relation of Eq.~(\ref{eq:barrio}) for selected modes. 
The parameters were 
$D=0.516$, $\alpha =0.899$, $\beta = -0.91$ and $\delta = 2$ for $k=0.45$ (dashed line), 
$D=0.122$, $\alpha =0.398$, $\beta = -0.4$ and $\delta = 2$ for $k=0.84$ (dash-dot line), 
$D=0.516$, $\alpha =0.89$, $\beta = -0.99$ and $\delta = .25$ for $k=0.96$ (solid line).
The region above $\lambda(k)=0$ bounds the wave number values for unstable modes.} 
\label{fig:dispersio}
\end{center}
\end{figure}

The onset of instability is  
subject to the following conditions: if $\alpha \ge 0$ then $\beta \le -\alpha$,
if  $\alpha \le 0$ then  $\beta \le -1$, and 
$\alpha -2 \sqrt{\alpha D} > \beta D$.
With these conditions, the wavenumber of the most unstable mode is given by
\begin{equation}
k_c^2 = \frac{D(\alpha-\beta)-(D+1) \sqrt{\alpha D}}{\delta D (D-1)}.
\label{eq:mostunstable}
\end{equation}
This equation was used to select parameters for the simulations. 
It is worth noticing that the algebraic forms of $B$ and $C$ in
Eq.~(\ref{eq:quadr}) make it difficult to analyze the system further. 
This is why Eq.~(\ref{eq:mostunstable}) is very useful. Fig.~\ref{fig:dispersio}
shows the real parts of the eigenvalues plotted against the wave number $k$ 
for the three modes we have used. 

\subsection{Gray-Scott model}

The second system we study is the Gray-Scott \cite{grayscott} model corresponding 
to two irreversible chemical reactions
\begin{eqnarray}
\mbox{U}+2 \mbox{V} & \rightarrow & 3\mbox{V} \nonumber \\
\mbox{V} & \rightarrow & \mbox{P},
\label{eq:chemical}
\end{eqnarray}
i.e., the reaction of chemical U with two parts of V produces three parts of V. Due to the
irreversible nature of the reactions, chemical P is an inert product. To write down the 
differential equations for the process, it is assumed that chemical U is fed in the reaction 
with constant rate $F$, the inert product is removed from the system, and that the second reaction 
is described by the rate constant $K$.
The equations of motion for the concentrations of the two chemicals $u \equiv u(\vec{x},t)$ and
$v \equiv v(\vec{x},t)$ in dimensionless units are as follows
\begin{eqnarray}
\partial_t u & = & D_u \nabla^2 u -uv^2  + F(1-u) \nonumber \\
\partial_t v & = & D_v \nabla^2 v + uv^2 - (F+K)v,
\label{eq:grayscott}
\end{eqnarray}
where it has been further assumed that chemical U is being consumed by the reaction at 
a rate that is proportional to $uv^2$. The  diffusion coefficients for the two chemicals
are $D_u$  and $D_v$, respectively.

The Gray-Scott model was studied analytically, and numerically in two dimensions, 
by Pearson \cite{xmorphia} who mapped the phase diagram for the system in terms of the
two rate constants for the reactions. 
This model exhibits a very rich behavior ranging from time-independent steady solutions
to chaotic, oscillatory, and to time-dependent phase turbulent behavior. Furthermore, 
Vastano {\it et al.} \cite{vastano} have shown that the system develops spatially steady patterns 
even when the two diffusion constants are equal. This behavior is particular to the 
one-dimensional case and it has not been observed in other dimensions. 

In this study, we first performed numerical simulations of the Gray-Scott model in three 
dimensions. After that, we randomly added a small number of sources for chemical V.  
The motivation for adding the sources was to investigate the robustness of patterns and the 
formation of connections between the sources, thus attempting to mimic the development 
of a biological network, e.g., the formation of synaptic contacts between neurons. 
Currently, it is unclear whether Turing systems can produce such an inductive 
signaling mechanism for neural patterning \cite{olivares,teemu,teemu2} but the idea is nevertheless
plausible.

\section{Numerical Calculations} 

First, we will present results from the numerical simulations 
of the general Turing system as given by Eq.~(\ref{eq:barrio}). 
Unless otherwise mentioned, all the simulations were performed in
a cubic domain of grid size $50 \times 50 \times 50$ using periodic 
boundary conditions. The well-known Euler algorithm was used for time integration 
with time step $dt=0.05$ for the general Turing system and 
$dt=1.0$ for the Gray-Scott model.

In addition, to test the robustness of the patterns, we performed 
a small number of simulations using Eq.~(\ref{eq:barrio}) 
with an added uncorrelated Gaussian noise source $\eta(\vec{x},t)$ 
with the first and second moments defined as 
$\langle \eta(\vec{x},t) \rangle = 0 $ 
and $\langle \eta(\vec{x},t)  \eta(\vec{x}',t') 
\rangle = 2 \varepsilon  \delta(\vec{x}-\vec{x}') \delta(t-t')$. 
The angular brackets denote an average and $\varepsilon$ 
is the intensity of the noise. 

\subsection{General Turing system}

In their study of the two-dimensional Turing system Barrio 
{\it et al.} \cite{barrio1} pointed out that the quadratic 
term in Eq.~(\ref{eq:barrio}) favors spots whereas the cubic
term enhances stripes. They also noticed that the spots were 
more robust. 

\begin{figure}[!]
\begin{center}
\includegraphics[width=.8\textwidth]{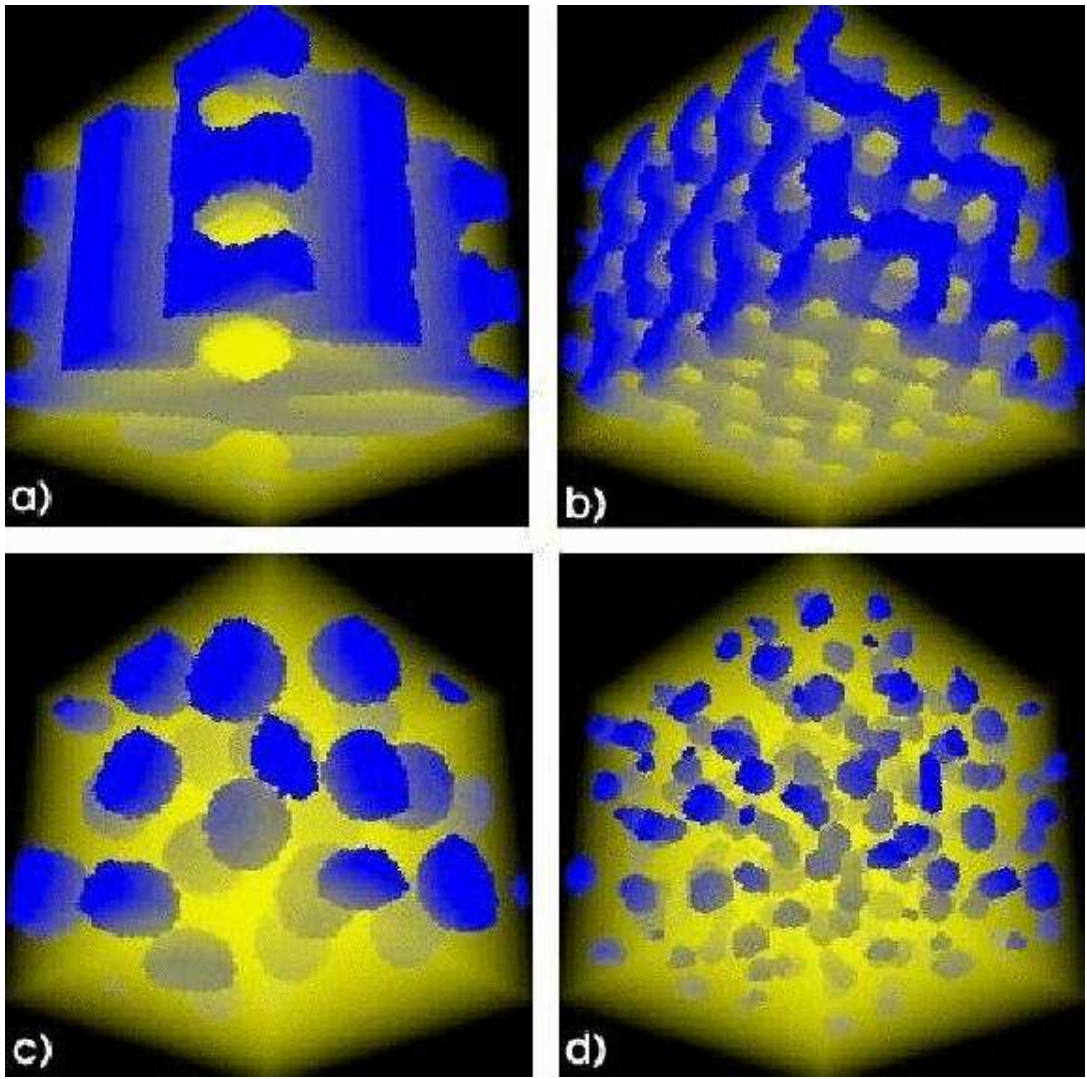} 
\caption{
Patterns obtained from simulations of the general Turing model
after $500~000$ iterations using random initial conditions
on every site of the lattice: $u_{0},v_{0} \in (0,1)$. 
Top row: $r_{1}=3.5$ and $r_{2}=0$; 
bottom row: $r_{1}=0.02$ and $r_{2}=0.2$.
Left column, the parameters were chosen to favor $k = 0.45$;
in right column to favor $k = 0.84$. 
} 
\label{fig:combo1}
\end{center}
\end{figure}

Fig.~\ref{fig:combo1} shows patterns obtained 
from simulations of the three-dimensional system. In the top row, 
the parameters were chosen to favor lamellar patterns, 
i.e., $r_2=0$. The figures imply that 
starting from completely random initial conditions, it is not 
likely for the system to converge into a purely lamellar state. 
This was confirmed by a number of extensive simulation runs.

The difference between  
Fig.~\ref{fig:combo1}a  and Fig.~\ref{fig:combo1}b is that the 
parameters $\alpha$ and $\beta$ have been selected in such a 
way as to enhance modes with wavenumbers $k=0.45$ and $k=0.84$, respectively. 
These are the most unstable modes in the sense of the linear 
approximation for parameter sets $\alpha=0.899$, $\beta = -0.91$, 
$\delta= 2$, $D=0.516$, and $\alpha=0.398$, $\beta = -0.40$, 
$\delta= 2$, and $D=0.122$, respectively. These particular choices 
of parameters were made in order to facilitate the comparison with the
2D results. The reason for these choices is that in the first
case, in the sense of the linear analysis, there are very few admissible modes
with positive growth rates and zero wavenumber degeneracy. As it is clear from 
Eq.~(\ref{eq:wavevector}), degeneracy becomes increasingly important 
at higher wavenumbers.
Ideally, the choice $k=0.45$ should produce patterns with 
wave vector $\vec{k}=2 \pi (n_x/L,0/L,0/L)$, i.e., favoring strongly 
the lamellar phase when $r_2 = 0$. However, as seen in 
Fig.~\ref{fig:combo1}a, the system has not reached a lamellar state
even after $500~000$ time steps but it has converged to a mixed
state instead. The simulations confirmed that to be the stable final 
state. 

Fig.~\ref{fig:combo1}b displays the situation where 
$k=0.84$ is favored. In this case, there are more closely 
spaced admissible modes leading to a competition between 
them.  It is important to notice that the system has a finite size 
and thus there are, in the sense of the linear analysis, only certain
admissible modes (Eq.~(\ref{eq:wavevector})). 

From the topmost figures of Fig.~\ref{fig:combo1} it is clear that the competition 
between the modes in the three-dimensional Turing system can lead to 
very interesting morphologies. In two dimensions, as studied 
by  Barrio {\it et al.} \cite{barrio1} (Figs. 2a and
2b therein), the corresponding patterns display stripes with 
a very small (or zero) number of topological defects. 
The defects can be considered as reminiscents of the more 
complicated pattern selection in 3D since
the 2D stripe patterns can be seen as cuts from a 
3D lamellar system. However, as seen in  
Figs.~\ref{fig:combo1}a and b, the 3D case displays much richer 
behavior and it is very difficult to obtain a purely
lamellar pattern starting from random initial conditions.  
It should also be observed that the difference between
Figs. 2a and 2b in \cite{barrio1} could be obtained by
varying only the wave length ($\delta$). However, in three
dimensions the qualitative difference of the two quantitatively 
different cases is clearer (Figs.~\ref{fig:combo1}a and b).

In the bottom row of Fig.~\ref{fig:combo1}, the parameters 
$r_1$ and $r_2$ have been selected to favor spots, or spherical
shapes (these can be compared directly to Figs. 2e and
2f in Barrio {\it et al.} \cite{barrio1}). The spherical patterns 
turned out to be very robust. In the simulations, these 
morphologies developed very fast and were also stable against 
random Gaussian fluctuations that were used to test the stability 
of the patterns against perturbations as discussed earlier. 

\begin{figure}[!]
\begin{center}
\includegraphics[width=.8\textwidth]{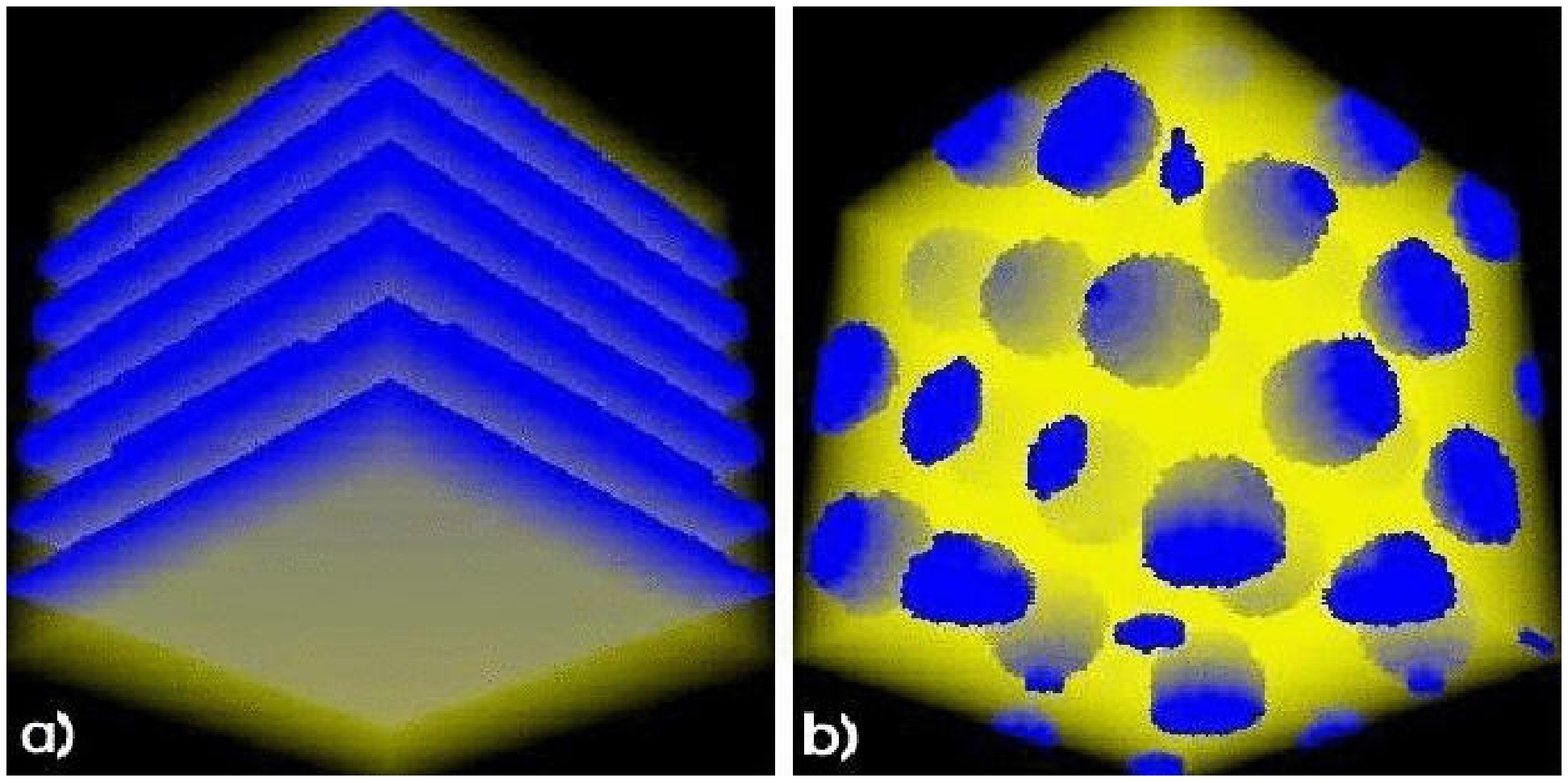} 
\caption{
Patterns obtained from simulation of the general Turing model
after $500~000$ iterations. Left: Initial conditions: chemical U
set in midplane, b) initial conditions: chemical U placed on 
111-plane in a triangular mesh.}
\label{fig:combo2}
\end{center}
\end{figure}

To study whether it is possible to obtain pure modes, we set up the system
in such a way that the initial conditions should favour only one selected mode.
Fig.~\ref{fig:combo2}a shows a situation where the system was prepared in such
a way that a layer of chemical U was set only in the midplane of the simulation box 
to provide favorable initial conditions for the lamellar structure to develop. 
The parameters are the same as used in Fig.~\ref{fig:combo1}b. In
Fig.~\ref{fig:combo2}b chemical U was put only in locations 
corresponding to the 111-plane in a triangular mesh, i.e., favoring hexagonal
symmetry of the spots obtained as in Fig.~\ref{fig:combo1}c. 
In both cases chemical V was initialized uniformly over the cube as in Fig.~\ref{fig:combo1}.
The stability of these patterns was also tested against Gaussian 
fluctuations and both of them turned out to be robust, although the planes
tended to align in a different way and curve slightly when a substantial amount of noise was used. 

\begin{figure}[!]
\begin{center}
\includegraphics[width=.8\textwidth]{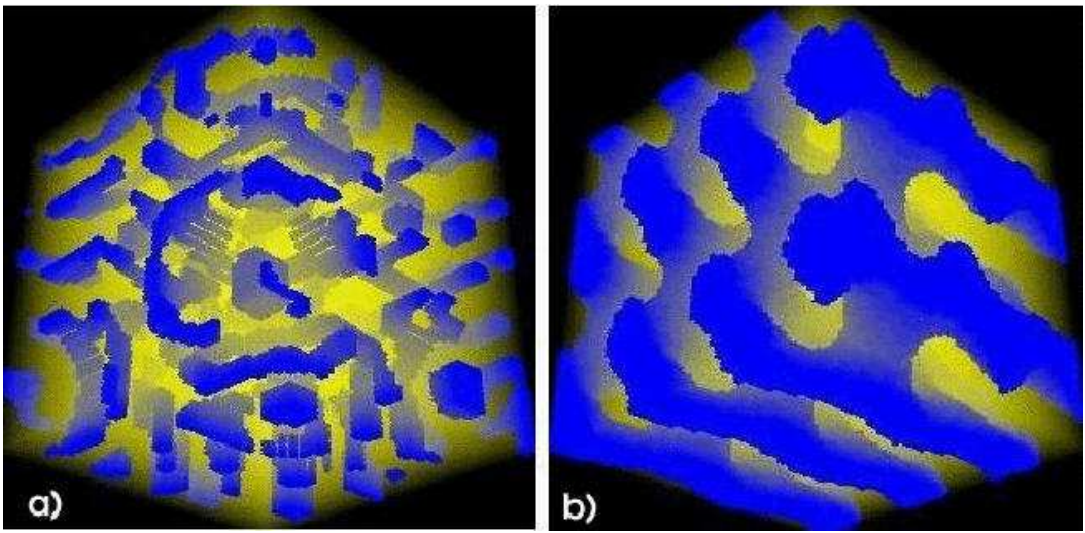} 
\caption{
Patterns obtained using Eq.~(\ref{eq:barrio}). 
a) After $500~000$ time steps. The isolated mode was $k=0.96$ and non-linear parameters 
$r_{1}=0.02$ and $r_{2}=0.2$, and 
b) after $2~000~000$ time steps, $k=0.45$, $r_{1}=3.5$ and $r_{2}=0.2$.
} 
\label{fig:combo3}
\end{center}
\end{figure}

To explore the effects of multiple modes on the morphology of the system,
we tuned the parameters in such a way that mode $k=0.96$ was the most unstable one.
In the simulations we used parameters $\alpha=0.89$, $\beta = -0.99$, $\delta = 0.25,
$ $D=0.516$. The non-linear parameters were $r_1 = 0.02$ and $r_2 = 0.2$ corresponding to spotty patterns. 
Fig.~\ref{fig:combo3}a shows a stabilized configuration where the pattern shows a tubular-like structure. 

To observe competition between spherical and lamellar structures we isolated mode $k=0.45$ and
set both $r_1$ and $r_{2}$ different from zero. Using non-linear parameters $r_1 = 3.5$ and $r_2 = 0.2$
resulted in the competition clearly seen in Fig.~\ref{fig:combo3}b. It took $2~000~000$ time steps 
to stabilize the structure.

\subsection{Gray-Scott model}

Finally, we performed simulations in two and three dimensions using the
Gray-Scott model of Eq.~(\ref{eq:grayscott}). In these simulations we have used 
$D_u = 0.125$ and $D_v = 0.05$ for the two diffusion constants. 
By scanning the phase space we obtained very rich behavior. However,
complexity and time dependence of some of the solutions caused
difficulties in visualising the resulting patterns in three dimensions.
The three-dimensional Gray-Scott model showed disordered lamellar-like and 
spotty-like phases. We also studied the case in which we distributed randomly four sources of 
chemical V in the system. The sources feed the chemical at a constant rate ($+0.01$). 
The shape of these sources is cross-like, having six branches in three dimensions
to x, y and z-directions, respectively.

\begin{figure}[!]
\begin{center}
\includegraphics[width=.8\textwidth]{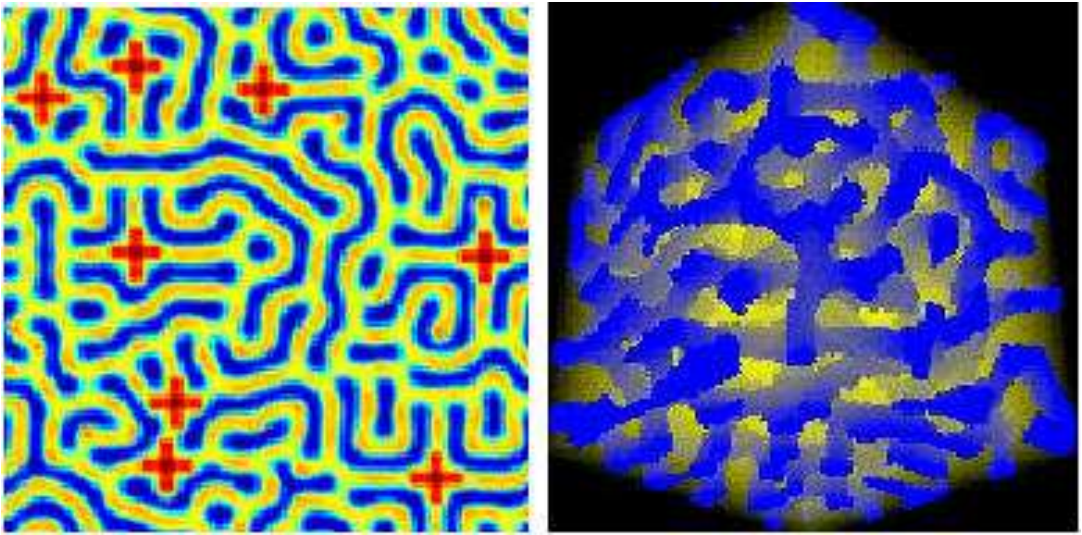} 
\caption{
a) Pattern obtained in a $120\times120$ lattice with periodic boundary 
conditions using the two-dimensional Gray-Scott model in
the presence of eight sources of morphogen V with parameters $F=0.065$, $K=0.0625$,
$D_{u} = 0.125$, $D_{v} = 0.05$. The sources appear as ``cross-like'' patterns.
b) Pattern obtained using the three-dimensional Gray-Scott model
with four sources of V, $F=0.045$, $K=0.065$, $D_{u} = 0.125$, $D_{v} = 0.05$.
} 
\label{fig:combo4}
\end{center}
\end{figure}

As discussed before, the motivation for including the sources is to investigate
if the Turing mechanism could be considered as a candidate for a mechanism
describing neuronal growth. Should that be the case, the sources can be thought 
of as representing neurons whereas the growing dendrites must connect them. 
In the sense of pattern formation in 3D systems, one requirement is the 
formation of stable tubular patterns. The Gray-Scott model clearly produces
them, see Fig.~\ref{fig:combo4}, and it is feasible to use it as the 
starting point to explore this topic further \cite{teemu2}.
The appearance of these tubular shapes seems characteristic for the
Gray-Scott model whereas it was difficult to obtain in the case of
the general Turing system. A systematic study is on the way to address
these issues \cite{teemu2}.

\section{Discussion}
In this paper, we have studied two reaction-diffusion models of Turing 
type in three dimensions. The first
model is the general model obtained by Taylor expansion and the other the 
well-known Gray-Scott model. 

We have shown that by enhancing more than one mode we 
can clearly see the competition between different modes. This is manifested as defects 
in the topology of the three-dimensional structures, whereas the patterns 
in two dimensions have very few defects. In turn this is due to the difficulty of
aligning planes compared with aligning lines in parallel. Also the spots
appear to be less regular in the case where there are more modes competing.
By favoring certain configurations with the choice of initial conditions we were able to drive
the system to a purely lamellar structure or hexagonal symmetry. We tested
the results against Gaussian noise and found the spots more robust in all
cases. To obtain chaotic behavior and competition we used wide window of values for the wave vector
and adjusted non-linear parameters.

The long range goal of our three-dimensional simulations is to model real biological
development, which can be done only in three dimensions. 
There are numerous examples of lamellar structures in nature, e.g. the skin and brain are formed out of layers.
As we showed the Gray-Scott model produces complex and stable tubular structures, where
the tubes show bifurcations.
In light of the results in two dimensions we believe that the branches connect the sources
also in three dimensions. 
It is known that neurons do not always form connections with
the nearest neurons, but with neurons far away, behind other cells. 
We believe that Turing patterns could explain this spatial selectivity by providing
the signaling pathways for neurotrophic factors. Simple diffusion of these substances 
cannot explain the complexity of neuronal patterning.

{\bf Acknowledgments.} 
We would like to thank Prof. P.K. Maini (Oxford, UK) for useful discussions
and Jarmo Pirhonen of the Centre for Scientific Computing (Espoo, Finland) 
for his help in visualization of the 3D Turing systems.
This work was supported by DGAPA UNAM through project IN104598
and by the Academy of Finland through its Centre of Excellence Programme.
One of us (RAB) thanks the Department of Theoretical Physics, Oxford
University of the hospitality during the preparation of this work.

\end{document}